\documentclass[fleqn,usenatbib]{mnras}

\usepackage{newtxtext,newtxmath}

\usepackage[T1]{fontenc}
\usepackage{ae,aecompl}

\usepackage{bm}
\usepackage[usenames, dvipsnames]{color}
\usepackage{subcaption}
\usepackage{graphicx}	
\usepackage{amsmath}	
\usepackage{amssymb}	

\graphicspath{{./Plots/}}

\title[Contamination in Catalogue Cross-Matching]{The Effect of Unresolved Contaminant Stars on the Cross-Matching of Photometric Catalogues}

\author[Tom J. Wilson and Tim Naylor]{
Tom J. Wilson,$^{1}$\thanks{E-mail: twilson@astro.ex.ac.uk}
and Tim Naylor$^{1}$
\\
$^{1}$School of Physics, University of Exeter, Stocker Road, Exeter EX4 4QL, UK\\
}

\date{Accepted 2017 March 10. Received 2017 March 10; in original form 2017 February 6}

\pubyear{2017}

\begin{document}
\label{firstpage}
\pagerange{\pageref{firstpage}--\pageref{lastpage}}
\maketitle

\begin{abstract}
A fundamental process in astrophysics is the matching of two photometric catalogues. It is crucial that the correct objects be paired, and that their photometry does not suffer from any spurious additional flux. We compare the positions of sources in WISE, IPHAS, 2MASS, and APASS with \textit{Gaia} DR1 astrometric positions. We find that the separations are described by a combination of a Gaussian distribution, wider than naively assumed based on their quoted uncertainties, and a large wing, which some authors ascribe to proper motions. We show that this is caused by flux contamination from blended stars not treated separately. We provide linear fits between the quoted Gaussian uncertainty and the core fit to the separation distributions. 

We show that at least one in three of the stars in the faint half of a given catalogue will suffer from flux contamination above the 1\% level when the density of catalogue objects per PSF area is above approximately 0.005. This has important implications for the creation of composite catalogues. It is important for any closest neighbour matches as there will be a given fraction of matches that are flux contaminated, while some matches will be missed due to significant astrometric perturbation by faint contaminants. In the case of probability-based matching, this contamination affects the probability density function of matches as a function of distance. This effect results in up to 50\% fewer counterparts being returned as matches, assuming Gaussian astrometric uncertainties for WISE-\textit{Gaia} matching in crowded Galactic plane regions, compared with a closest neighbour match.
\end{abstract}




\section{Introduction}
\label{sec:intro}
Broadband photometry is a staple of astrophysics, able to provide a wealth of information on a plethora of objects of interest without the time requirements of spectroscopy. To break degeneracies in theoretical models and gain as much understanding as possible, oftentimes multi-wavelength coverage is required. This means combining the efforts of several surveys, where teams and collaborations have independently taken photometric images of the sky in various wavelength regimes. It is therefore of vital import that we correctly identify the same stars in separate catalogues.

Traditionally, the method for matching two catalogues together uses the smallest distance between a given star in one catalogue and stars in the opposing catalogue, pairing those stars that both have the other star as their closest corresponding star. Additionally there is a cutoff radius beyond which no pairs can be matched, typically 2" or 3". Due to the nearest-neighbour nature of the matching, this is typically referred to as proximity matching.

Recently, the idea of matching between catalogues following a probabilistic approach (starting with \citealp{Sutherland:1992aa}; more recently, e.g., \citealp{Budavari:2008aa}, \citealp{Fleuren:2012aa}, \citealp{Naylor:2013aa}, and \citealp{2017PASA...34....3L}) has become common. Instead of merely assigning a maximum match radius, the methods calculate the probability of finding a star's counterpart in a second catalogue at a given separation. These probabilities are based on the uncertainty in the position of the star in each catalogue. This we will refer to as probability-based matching. It gives a more flexible approach by adjusting the size scale over which matches are considered likely to match the precision of the detections. High quality, precise astrometric data only allow matches between stars close to one another, while less precise data are allowed to have counterparts beyond the 2-3" typical proximity cutoff.

Proximity matching is equivalent to carrying out probability-based matching using a ``top-hat'' function with the cutoff radius, inside which a star is equally likely to exist at any distance from another detection and outside which it is impossible to be matched. Astrometrically the full probability-based method is favourable because the top-hat is unphysical. To improve upon this ``top-hat'', we require a more complete description of the probability of detecting the counterpart in the opposing catalogue at a given separation. These probabilities of star pairs being counterparts to one another as a function of separation are themselves a function of what we shall refer to as the astrometric uncertainty functions (AUFs). Usually, these distributions are assumed to be purely Gaussian. This does not account for any wings to the distributions themselves, yet these are known to exist (see, e.g., \citealp{Krawczyk:2013aa} Figure 4 or \citealp{Munari:2014aa} Figure 2). The assumption that the AUF is Gaussian could lead to a significant mis-identification of a large number of counterparts. In the probability-based matching case this incorrect matching is due to the assumed shape of the distributions not being a good description. In the proximity matching case it is caused by the accepted cutoff radius being too small.

Probability-based matching also has increased flexibility in allowing for comparisons between two detections in one catalogue by including additional information, such as magnitudes (e.g., \citealp{Budavari:2008aa} and \citealp{Naylor:2013aa}). If two stars are close enough to the same star in another catalogue to be considered likely matches, the extra parameter space allows for the possibility of rejecting an unfavourable match that is serendipitously nearer than the better match. However, this extra information can not be used if the AUFs are ill-defined, so it is vital that they are correct.

In this paper we will explain how crowding in high density regions causes long, non-Gaussian tails in the AUFs. We will begin by initially introducing the catalogues being used throughout the paper in Section \ref{sec:catalogues}, and in Section \ref{sec:distributions} defining the AUF more formally. We will then examine the spatial distribution of an examples of matches for a crowded region of the Galactic plane before discussing some possible reasons for the non-Gaussianity seen in the distributions, concluding that they cannot satisfactorily explain the results in Section \ref{sec:fittingthedist}. We introduce the effect of crowding seen in photometric catalogues in Section \ref{sec:fitting}. This is used to explain how this effect causes the non-Gaussian tails, before we test the hypothesis with some simple approximations in Sections \ref{sec:synthetictests} and \ref{sec:contamlevels}. We then put the effect into context for several additional large scale, commonly used surveys in Section \ref{sec:surveycontext}. Finally, we offer some options to overcome the issue of contamination in Section \ref{sec:summary}. Here we give some cases where one can maximise the number of true matches at the expense of false positives, or, alternatively, minimise the number of false positives and contaminated matches. We define symbols used throughout the paper in Table \ref{tab:symbols}.

\section{Catalogues}
\label{sec:catalogues}
The matching of photometric catalogues has significant problems in very crowded fields, and is at its worst in the Galactic plane, especially towards the Galactic centre. In addition, the crowding becomes more problematic with increasing seeing or larger point spread functions (PSFs). The crowding of stellar fields is then a function of both stellar density and PSF area, which is why we have chosen to focus on WISE \citep{Wright:2010aa} for most of our work. With a $\simeq6"$ full-width at half maximum (FWHM) in bands $W1-W3$ and a relatively deep survey reaching $W1 \simeq 17$, the WISE datset suffers from significant crowding. At the other extreme, the recently released {\it Gaia DR1} \citep{Gaia-Collaboration:2016aa} provides excellent and unprecedented astrometric precision, and with a $\simeq0.1"$ FWHM should be effectively uncrowded.

Initially we will consider \textit{Gaia} and WISE, but we will introduce APASS \citep{Henden:2014aa}, IPHAS \citep{Barentsen:2014tb}, and 2MASS \citep{Skrutskie:2006um} in a later Section. To ensure minimal erroneous or poor data in the catalogues, we first clean them to remove either known non-stellar sources, or to remove spurious, low-quality, saturated, and upper flux limit objects, as detailed in Table \ref{tab:flagtable}.

\begin{table}
\begin{tabular}{c | l}
\hline
Symbol & Definition \\
\hline
$A$ & Area\\
$F$ & Flux ratio of bright and faint objects\\
$g$ & Probability density of matches at given distance\\
$l$, $b$ & Galactic sky coordinates\\
$M$ & Total number of counterparts\\
$m$ & Magnitude difference between faint and bright object\\
$m_0$ & Magnitude of bright object\\
$N$ & Number of stars per unit area per magnitude\\
$r$ & Radial distance\\
$Q$ & Contamination figure of merit \\
$R$ & Cutoff radius\\
RA, Dec & Celestial coordinates\\
$U, V$ & Number of objects in circle of given radius\\
$x$, $y$ & Cartesian coordinates\\
$z$ & Scaling for increase in star counts with magnitude\\
$\Delta r$ & Width of radial annulus\\
$\mu_\mathrm{RA}$, $\mu_\mathrm{Dec}$ & Proper motion in sky coordinates\\
$\sigma$ & Astrometric Gaussian uncertainty \\
$\sigma_\mathrm{quoted}$ & Astrometric uncertainty given in catalogue\\
$\sigma_\mathrm{core}$ & Uncertainty fit to the inner radius of an AUF\\
$\frac{\mathrm{d}N}{\mathrm{d}r}$ & Number of separations per unit distance\\
$\frac{\mathrm{d}N}{\mathrm{d}A}$ & Number of stars per unit area\\
$\frac{\mathrm{d}N}{\mathrm{d}A}_\mathrm{cat}$ & Number of detected sources per unit area\\
\hline
\end{tabular}
\caption{Table showing the definition of symbols used throughout.}
\label{tab:symbols}
\end{table}

\begin{table*}
\centering
\begin{tabular}{c | c}
\hline
Catalogue & Criteria\\
\hline
\textit{Gaia} & astrometric\_excess\_noise $>$ 0.865mas; or matched\_observations $\leq$ 8 or\\
 &  astrometric\_n\_good\_obs\_al + astrometric\_n\_good\_obs\_ac $<$ 60\\
\hline
WISE & ``Contam'' flag is either ``D'', ``P'', ``H'', or ``O''; or ``ext'' flag is 2, 3, 4, or 5; or\\
 & ``Phqual'' flag is ``X'' or ``Z''; ``detbit'' == 0; Mag == NaN; ``sat'' flag $>$ 0; or $\sigma_{\mathrm{Mag}}$ == NaN\\
\hline
APASS & Mag $>$ 20 or Mag $<$ 10\\
 \hline
2MASS & ``Galcontam'' or ``Mpflag'' flags set; or ``Blend'' flag == 0; ``Read'' flag == 0 or 3; Mag == NaN; or $\sigma_{\mathrm{Mag}}$ == NaN\\
\hline
IPHAS & $p_{\mathrm{star}} < 0.9$; or Mag == NaN, ``Saturated'' flag set, or $\sigma_{\mathrm{Mag}}$ == NaN\\
 \hline
\end{tabular}
\caption{Table showing the various flags for rejection from the catalogues used.}
\label{tab:flagtable}
\end{table*}

\section{The Astrometric Uncertainty Function}
\label{sec:distributions}
The probability that two stars in two photometric catalogues are counterparts to one another is the probability that the stars from the two catalogues are drawn from the same original sky position, involving the AUFs of both catalogues. However, the order-of-magnitude higher precision in the \textit{Gaia} dataset simplifies the problem such that the probability of matches reflects only the uncertainties in the second catalogue. Thus, we only require the AUF of WISE detections in this instance.

This means we can model the probability of measuring a source, with ``true'' position at the origin, at position $x, y$ as a centered, circular, two-dimensional Gaussian (Quetelet, summarised by \citealp{Herschel:1857aa})
\begin{align}
\begin{split}
g(x, y, \sigma) &= \frac{1}{2\pi\sigma^2}e^{-\frac{x^2 + y^2}{2\sigma^2}},
\label{eq:bivariatepdf}
\end{split}
\end{align}
where $\sigma$ is the astrometric uncertainty in either of the orthogonal axis directions. The astrometric uncertainty can be approximately related to the photometric signal-to-noise ratio (SNR) and image PSF scale length. \citet{King:1983aa} quotes the relationship as the FWHM of the image divided by the SNR.

When considering a circular geometry, we can transform this to radial coordinates by integrating over $\theta$, which changes the Gaussian distribution to a Rayleigh distribution, given by

\begin{align}
\begin{split}
g(r, \sigma) &= \frac{r}{\sigma^2}e^{-\frac{r^2}{2\sigma^2}}.
\label{eq:rayleighdist}
\end{split}
\end{align}

$g(x, y, \sigma)$ is a probability density function, the probability per unit area, that the WISE star will be detected at an offset $x,y$ from the \textit{Gaia} source. Alternatively, $g(r, \sigma)$ is the probability per unit length that the WISE star is detected at a radial offset $r$ from the \textit{Gaia} source. It is the function $g(r, \sigma)$ that we will compare to our data in Section \ref{sec:fittingthedist}.

\section{Fitting the Distribution}
\label{sec:fittingthedist}
To check the validity of $g$, our AUF, we must test it against some example data. Consider a large sample of matches, i.e. pairs of stars, all of which have a similar astrometric uncertainty $\sigma$. The number of matches per unit distance in a narrow annulus $r$ to $r + \Delta r$ is
\begin{align}
\begin{split}
\frac{\mathrm{d}N}{\mathrm{d}r}(r, \sigma) &= \frac{M}{\Delta r}\int\limits_r^{r + \Delta r}\!\frac{r}{\sigma^2}e^{-\frac{r^2}{2\sigma^2}}\,\mathrm{d}r,
\label{eq:numofmatches}
\end{split}
\end{align}
where $M$ is the total number of matches. Assuming all stars in the sample are true matches (see Section \ref{sec:commonmatches} for further discussion), we can then compare our expected number of stars per unit distance with the number detected. 

In this section we will consider matches between WISE and the Tycho-Gaia Astrometric Solution (TGAS; \citealp{2015A&A...574A.115M}) for an 800 square degree region of the Galactic plane ($100 \leq l \leq 140$, $-10 \leq b \leq 10$). Although the TGAS is a relatively bright subset of the full \textit{Gaia} dataset, limiting our match numbers, we will require the proper motions, which are only available for TGAS stars, in Section \ref{sec:commonmatches}. We will discuss the effects of the full magnitude range in Section \ref{sec:surveycontext}, and find the magnitude cut does not affect the conclusions drawn in this Section.

\subsection{Uncertainties for WISE Data}
\label{sec:uncertsforwise}
\begin{figure}
    \centering
    \includegraphics[width=\columnwidth]{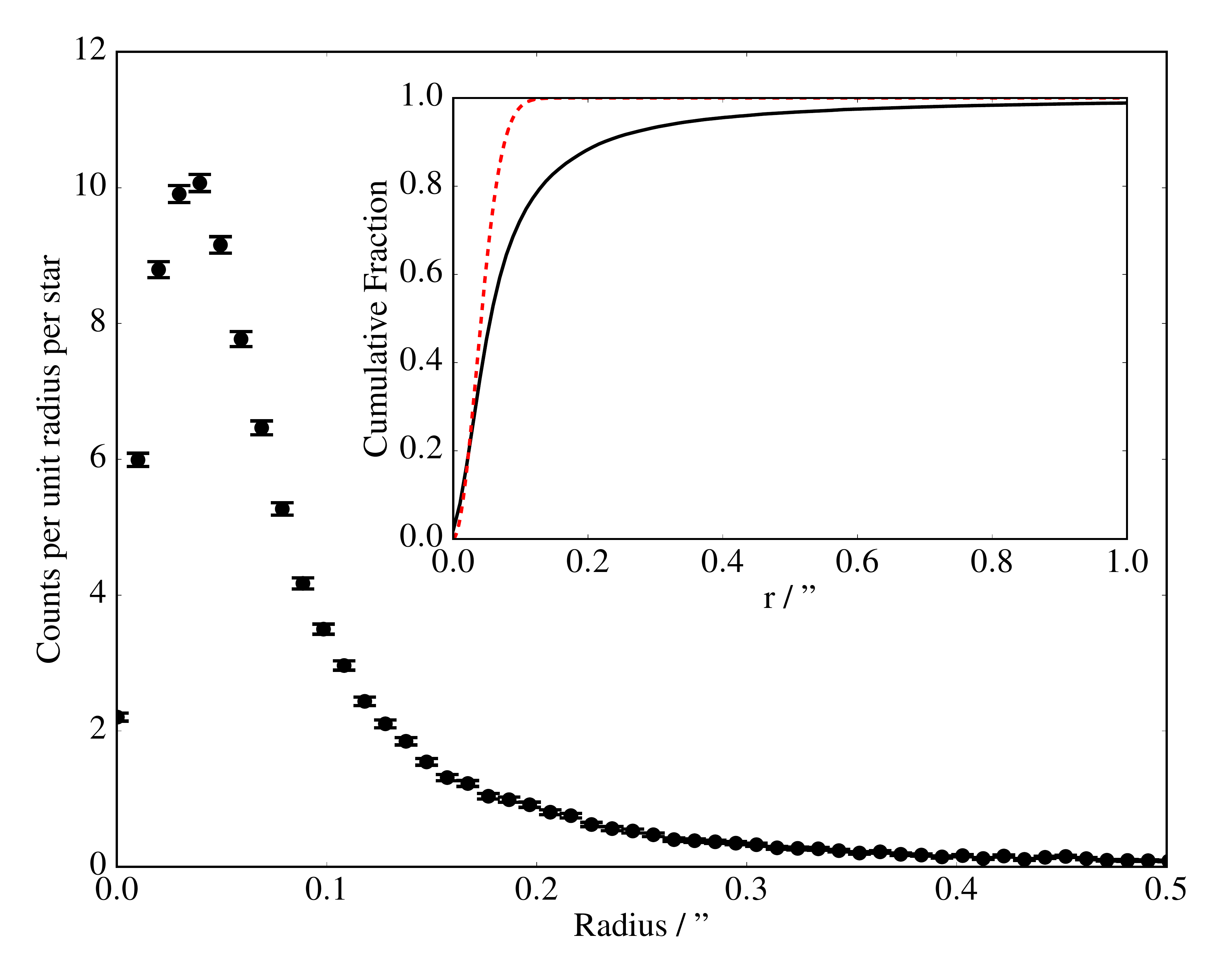}
    \caption{Figure showing separation of proximity matches between TGAS and WISE, for WISE objects with quoted uncertainty $\sigma = 0.039\pm0.001"$. Inset Figure shows the cumulative distribution, with reference cumulative Rayleigh distribution of $\sigma = 0.039"$ shown as a red dashed line.}
    \label{fig:brightplane}
\end{figure}

Matching between our two catalogues, we take WISE stars in a narrow range of $\sigma$ values (typically $\lesssim$ 0.01") and proximity match them in a nearest-neighbour scheme to the TGAS dataset. From this we find the number of stars in given radius bins, and plot the number of stars per unit radius within each annulus, along with the assumed astrometric distribution, based on the quoted uncertainties. Figure \ref{fig:brightplane} shows the resulting distribution for one narrow range of uncertainties $\sigma = 0.039\pm0.001"$. We can see that the distribution is reasonably well described by a Rayleigh distribution in the inner region, below $r \simeq 0.1"$, but that there is a significant non-Gaussian tail to the distribution of match distances.

\subsection{Common Sources of Additional Astrometric Sources}
\label{sec:commonmatches}
There are two obvious potential causes of non-Gaussian data: a population of uncorrelated false matches, and the effects of proper motion on the apparent match distance between two catalogues of different epochs. As we show below neither of them can adequately explain the effect entirely, requiring an alternative explanation.

\subsubsection{Proper motions}
\label{sec:propermotions}
Proper motions are often cited as being the cause of these ``wings'' at large separations (e.g., Section 6.4 Figure 2 of \citealp{Cutri:2012aa}, Appendix A1 of \citealp{Flesch:2004aa}). As WISE operated in 2010 while \textit{Gaia} {records positions in epoch J2015} we must check to see if this is a significant cause of match offsets. We obtained the \textit{Gaia} proper motions in the orthogonal axes for all stars in the 800 square degree region of the Galactic plane used to construct the distributions in Figure \ref{fig:brightplane}.

\begin{figure}
    \centering
    \includegraphics[width=\columnwidth]{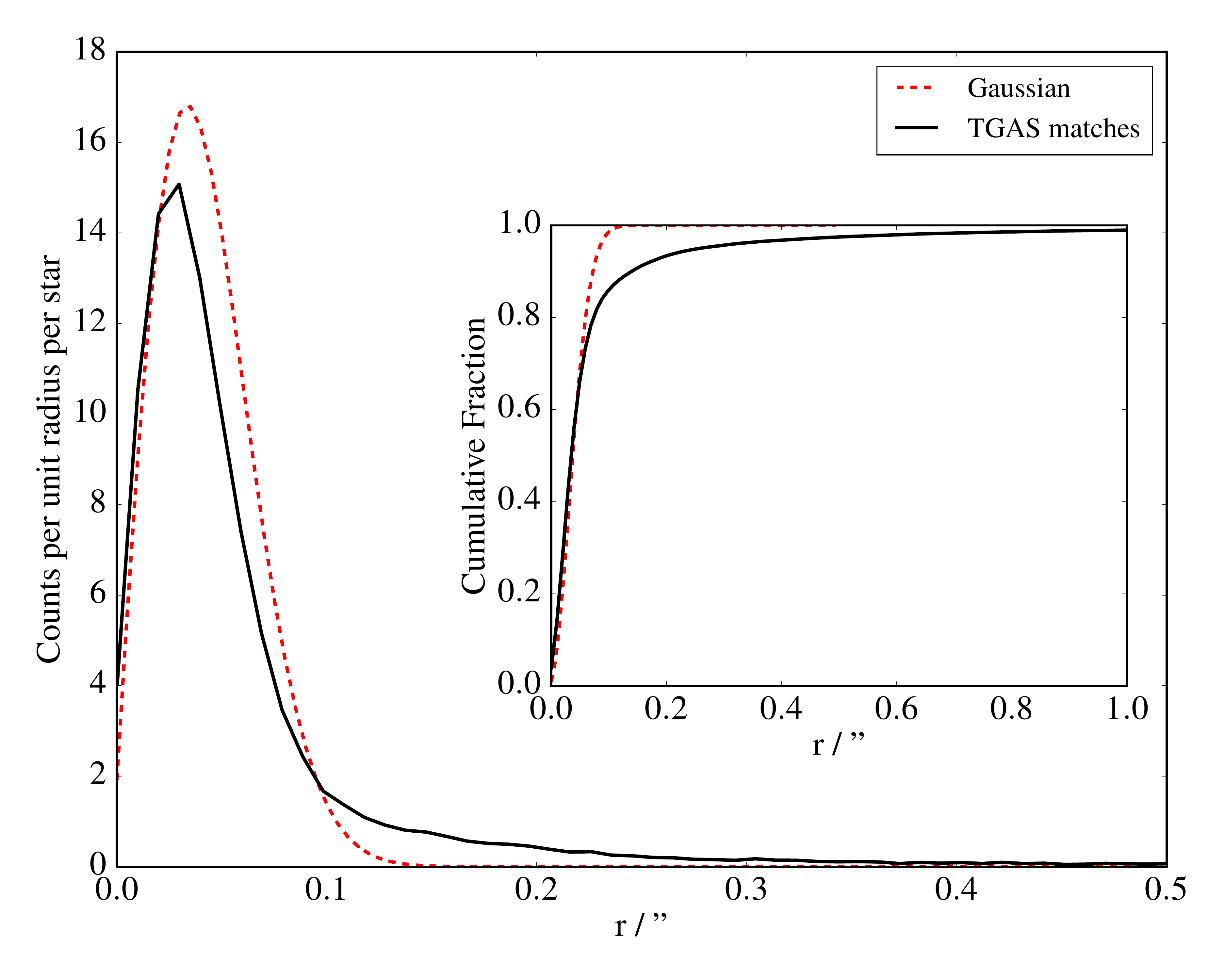}
    \caption{The effects of proper motions on WISE-TGAS matches with WISE astrometric uncertainty $\sigma = 0.039\pm0.001"$. The distribution of separations, corrected for proper motion during the five year gap between observations, is shown as a solid black line. These are compared to the expected Gaussian of uncertainty $\sigma=0.039"$, shown in the red dashed line. The proper motion correction fails to account for most of the matches seen at large separations in the non-Gaussian tail.}  
    \label{fig:W1brightPMTGAScounts}
\end{figure}

We calculated the new celestial coordinates for the \textit{Gaia} positions, transformed from the J2015 epoch to WISE's J2010 epoch as

\begin{equation}
\mathrm{RA}_\mathrm{new} = \mathrm{RA} - 5 \mathrm{year} \cdot \mu_\mathrm{RA}\left[\cos(\mathrm{Dec})\right]^{-1},
\label{eq:pmdrift}
\end{equation}
with an equivalent transformation for declination, where $\mu_\mathrm{RA}$ and $\mu_\mathrm{Dec}$ are the projected proper motions in the two orthogonal sky axis directions. The new distribution of proper motion-corrected separations was compared to a Gaussian of the average uncertainty $\sigma=0.039"$, shown in Figure \ref{fig:W1brightPMTGAScounts}. As can be seen, while the distribution tightens slightly towards smaller separations, the large, non-Gaussian tail remains beyond $r \simeq 0.1"$. This leads to an incompatible cumulative distribution shown inset to Figure \ref{fig:W1brightPMTGAScounts}. The non-Gaussian tail increases with decreasing brightness (see Section \ref{sec:contamlevels} for more details), and the average magnitude of stars in Figure \ref{fig:W1brightPMTGAScounts} is bright, at W1$\simeq11$. We therefore cannot explain most of the non-Gaussianity of the distributions with proper motions.  

\subsubsection{Uncorrelated False Matches}
\label{sec:falsematches}
While we cannot explain the non-Gaussianity to the match distributions with proper motions, these are purely proximity matches. We expect some contamination from uncorrelated stars which could potentially explain the non-Gaussian wings. At its most dense, there are $2\times10^4$ \textit{Gaia} stars per square degree in the Galactic plane region in question. The expected number of randomly placed objects in a circle of a given radius, $U$, is the multiple of the stellar density, $\frac{\mathrm{d}N}{\mathrm{d}A}$, and the area, $A$,

\begin{equation}
U = \frac{\mathrm{d}N}{\mathrm{d}A} \times A = 2\times10^4\,\mathrm{deg}^{-2} \times \pi \left(\frac{0.5"}{3600"/\mathrm{deg}}\right)^2 = 0.0012,
\label{eq:expected3"contam}
\end{equation}
where we have limited ourselves to a circle of radius 0.5" as per Figure \ref{fig:brightplane}. We therefore expect 0.1\% of the stars to be false matches. These numbers are upper limits, as the nearest-neighbour scheme employed reduces contamination beyond the radius of the true match separation for each star. We conclude that we cannot explain the distribution wings with uncorrelated star contamination. 

\section{Explaining the Distribution Wings}
\label{sec:fitting}

\subsection{Star Spatial Distributions}
\label{sec:spatialstardists}
\begin{figure}
    \centering
    \includegraphics[width=\columnwidth]{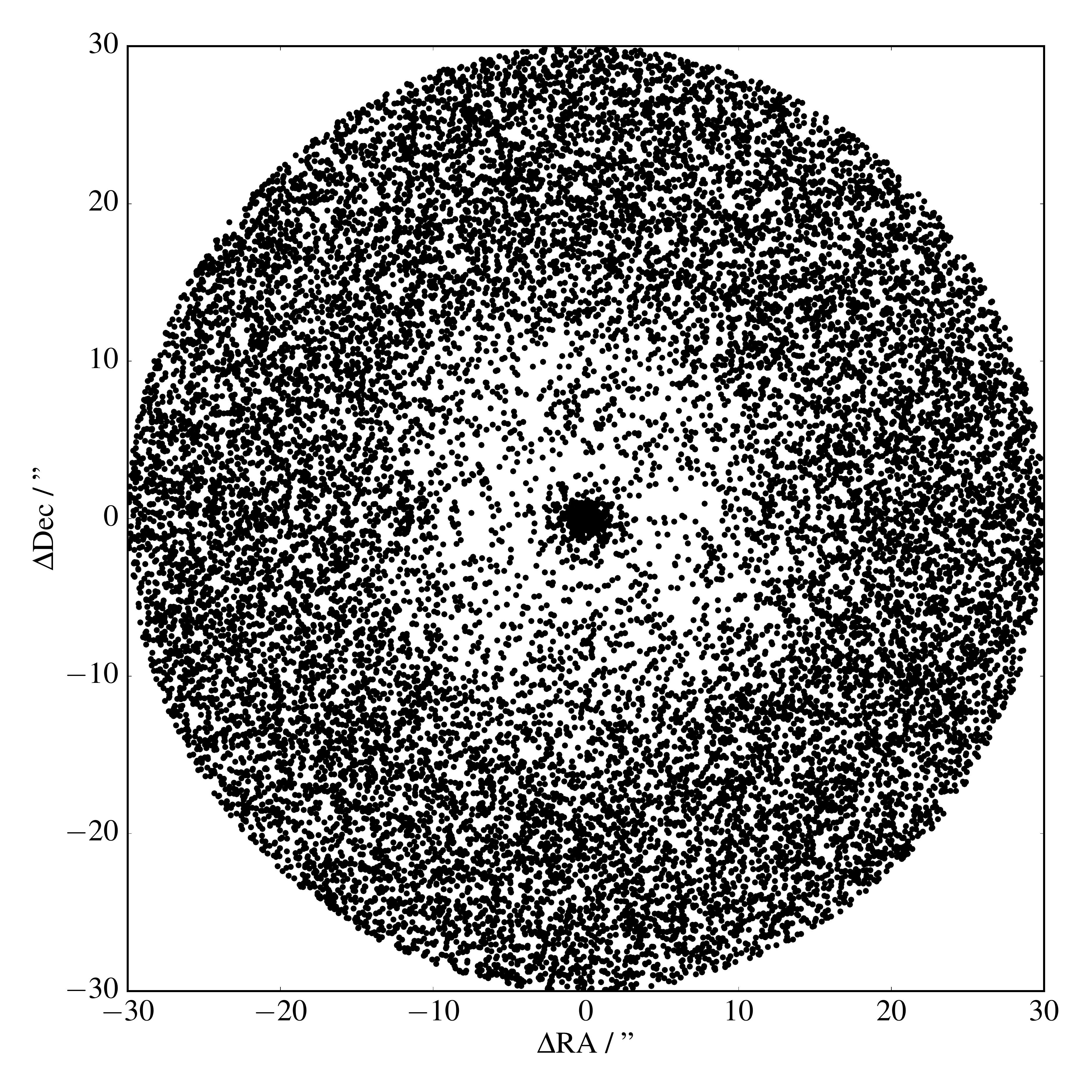}
    \caption{Figure showing the spatial separation of all WISE stars within 30" of \textit{Gaia} sources $15 \leq G \leq 15.25$, for a 5$^\circ \times 5^\circ$ slice of the Galactic plane. Background sources are seen at a constant density surrounding a clump of counterpart stars in the centre. However, the background density decreases within $\lesssim 10"$ due to the crowding out of the fainter background sources by bright counterparts.}
    \label{fig:crowding}
\end{figure}

To explain the distribution of matches between two catalogues, it is illuminating to consider a \textit{Gaia} source of magnitude $15 \leq G \leq 15.25$. We can find the offsets from this star to all WISE objects with radial offset <30". Repeating this calculation for all such stars in a 25 square degree region of the Galactic plane at $120 \leq l \leq 125$, $0 \leq b \leq 5$ we build up a density of WISE sources astrometrically near \textit{Gaia} sources in a narrow \textit{Gaia} magnitude range as a function of radial distance, shown in Figure \ref{fig:crowding}. 

There are three distinct regions. First, beyond 10" from the \textit{Gaia} objects we have a constant density of sources, which are uncorrelated, additional WISE objects. Second, we have a tight clustering of detections inside $r \lesssim 2"$, which are the WISE detections corresponding to our \textit{Gaia} objects. Third, we have a region $2" \lesssim r \lesssim 10"$ where we see randomly placed objects at a lower density than those at larger $r$.

However, non-match stars - those in the WISE catalogue whose $G$ magnitude would lie outside of our 0.25 magnitude range - are not correlated with those stars that do lie in that small magnitude range. We therefore expect them to have a constant stellar density across the entire sky, meaning that between 2" and 10" radial distance we should see the same density of objects in some small area as we do beyond 10". This apparent reduction in stellar density is caused by crowding, a well known issue where bright sources dominate and cause non-detections of fainter objects inside their PSF, reducing the number of objects measured at these intermediate distances. 

The important point to stress here is that these stars have not gone away - they are merely absorbed into the PSF of the bright star. This causes flux contamination, which will compromise the photometry. However, since the vast majority of the contaminating sources will be objects significantly fainter than the main detection, with a low relative flux ratio, the photometric effect is small. 

\subsection{Contaminant Stars}
\label{sec:contaminants}
\begin{figure}
    \centering
    \includegraphics[width=\columnwidth]{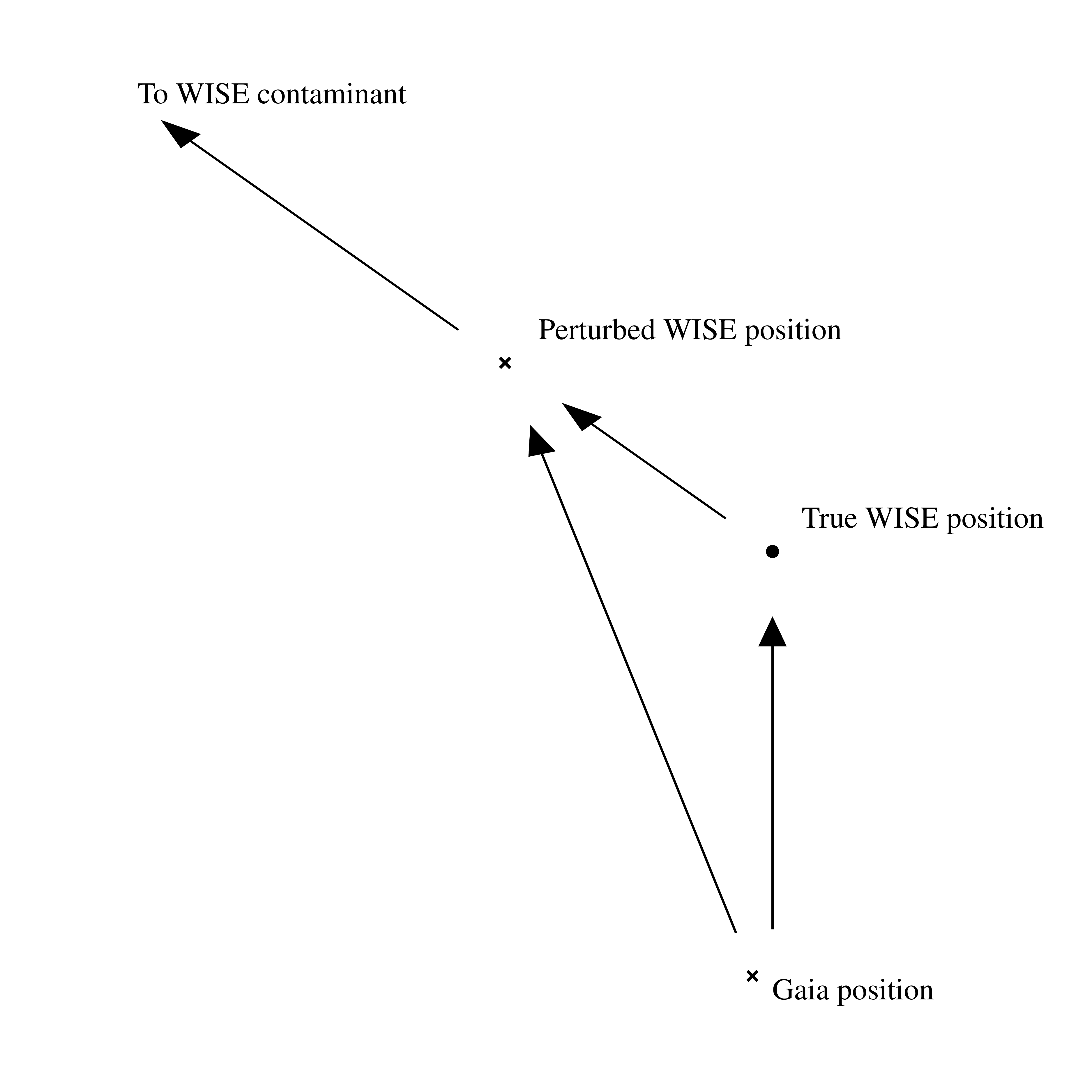}
    \caption{Figure showing the effect of unresolved contamination on the measured position. Here, a \textit{Gaia} object is separated from its true WISE counterpart by some distance. An undetected second WISE star within the WISE PSF causes the measured position to be shifted, causing a different separation to be calculated. This leads to a distribution of separations that is not merely based on Gaussian statistics.}
    \label{fig:shiftsschematic}
\end{figure}

More crucial, however, is the effect these sources have on the derived positions. Figure \ref{fig:shiftsschematic} shows an example schematic. A \textit{Gaia} source and its true WISE match are offset by some small distance - on the order of tenths of arcseconds - but there lies inside the $\simeq10"$ WISE PSF a second, undetected source with a tenth of the flux of the primary source, at $\simeq3"$. This will tug on the position of the WISE primary by 0.3", changing the apparent separation between the WISE object(s) and the \textit{Gaia} object. The distribution of separations - which we would wish to use for any probabilistic catalogue matching - is then a combination of two functions: the initial Gaussian-based statistics and the effects of undetected, embedded, contaminants.

\section{Validation with Synthetic Distributions}
\label{sec:synthetictests}
To test the effect these embedded stars could have on the AUF, we created a synthetic dataset based on simple geometric arguments. First we require the distribution of shifts that result when stars are contaminated within their PSF. 

To obtain the shift distribution, we placed test stars inside $10^5$ circles of a given sample bright star's PSF at random. These drawings assumed that the number density of stars increases by a factor of $z=2$ with every step in magnitude. We then found the flux-weighted position of the stars in each PSF. Once all test contaminants had been drawn, the number of new positions in each given distance bin was recorded. Finally, the distribution was reduced to a probability density function by normalising.
\begin{figure}
    \centering
    \includegraphics[width=\columnwidth]{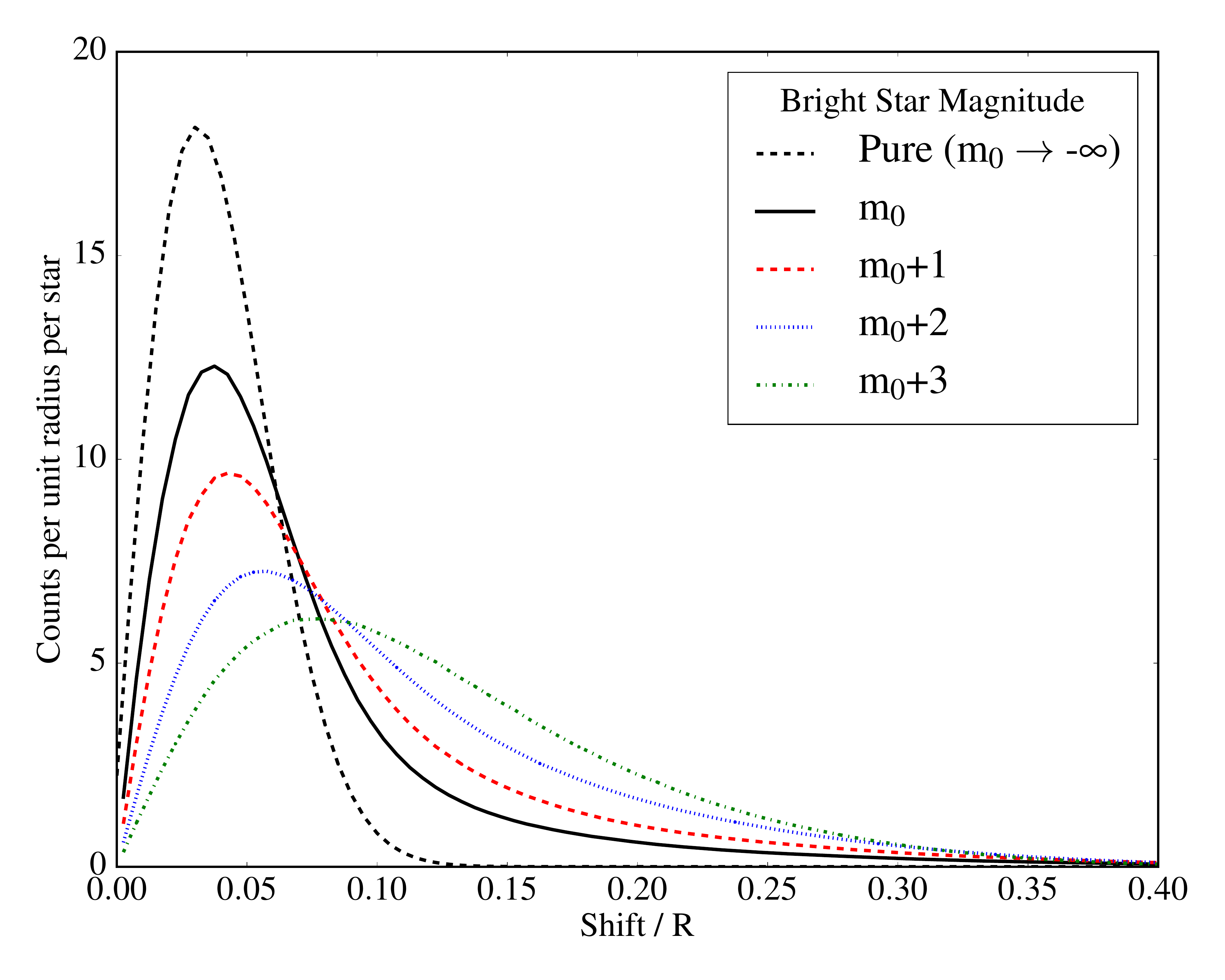}
    \caption{Figure showing the effect of unresolved contaminating stars on distributions of synthetic positions in units of the PSF cutoff radius $R$. A Rayleigh distribution with $\sigma = 0.05\times$FWHM was convolved with a derived contamination shifts distribution. The result is an inflation of the Rayleigh distribution uncertainty, as well as the introduction of the large, non-Gaussian tails similar to those seen in Figure \ref{fig:brightplane}, increasing with increasing stellar number density. The magnitudes, $m_0$ through $m_0+3$, represent increasing magnitudes of the central bright star, with a corresponding increase to the number densities of contaminants. The pure Rayleigh distribution, which effectively represents the contamination effects on an infinitely bright central star, is also plotted for reference.}
    \label{fig:fakeshifts}
\end{figure}

We convolved the resultant function with a Rayleigh distribution of $\sigma = 0.05\times$FWHM, representing a star with SNR=20. The results of this are shown in Figure \ref{fig:fakeshifts}, for several bright stars with increasing magnitudes, representing increasing number densities of sky objects. The convolved functions still resemble the ``pure'' AUF in the inner region of the PSF, albeit with a broadened equivalent astrometric uncertainty, but the contamination also introduces a very long tail of separations. These objects are flux contaminated enough to introduce offsets on the order of 0.3-0.4$\times$FWHM. This effect increases as the number density of objects increases, representing increased large separation contamination.\\

In summary, we suggest that the effect of astrometrically perturbed sources leading to large wings in distributions of counterpart distances, seen in the number of astrometric separations as a function of distance, is caused by the crowding out of fainter objects in the PSF. This leads to that fraction of stars - a very large fraction in regions of high stellar density, faint magnitudes, or large PSFs - with contaminant stars buried in their PSF exhibiting significantly non-Gaussian distributions in their detected positions. This will cause additional missed proximity matches if using a cutoff radius on the order of 1-2". It will also cause the resultant likelihoods derived from any probabilistic catalogue matching methods to fail in sampling the correct probability of matches and non-matches, also leading to a large fraction of false negative assignments.

\section{Quantifying the Contamination Levels}
\label{sec:contamlevels}
We showed that simple arguments about the effects of faint embedded stars inside brighter PSFs can reproduce similar results to those seen in the data (Figure \ref{fig:fakeshifts} cf. Figure \ref{fig:brightplane}) in Section \ref{sec:synthetictests}. However, we must now quantify the contamination levels from those faint stars. At a given stellar magnitude there will be some fraction of stars containing unresolved stars and another fraction which do not have within them additional sources. These are contaminated and uncontaminated objects, respectively. The uncontaminated fraction will still obey traditional Gaussian-based probabilistic statistics, but the contaminated stars will exhibit large shifts to their apparent position. This leads to the significant wings in their AUFs, as seen in, e.g, Figure \ref{fig:brightplane}.

The average number of stars inside the circle of the PSF can be calculated in a similar way to Equation \ref{eq:expected3"contam}, but for the area we use the radius of the circle the PSF subtends on the sky - typically 1-1.5 times the FWHM. In addition the number density is now the number of stars per square degree up to $m$ magnitudes fainter than the star of magnitude $m_0$. This then gives us a fraction of stars which are contaminated,

\begin{equation}
V = \frac{\mathrm{d}N}{\mathrm{d}A} \times A = \int\limits_{m_0}^{m_0 + m}\!N z^m\,\mathrm{d}m \times \pi R^2,
\label{eq:contaminantfraction}
\end{equation}
with $R$ the PSF cutoff radius, $N$ a normalisation factor, and $z \simeq 2$ the increase in stellar density with each step in magnitude. The choice of $m$ is a reasonably arbitrary one, with stars technically being contaminated by faint stars with vanishingly small flux ratios, requiring an upper limit to the integral approaching infinity. However, the test data used in Section \ref{sec:synthetictests} show a convergence of the resulting AUFs for $m \gtrsim 4$. This suggests that above $m\simeq4$, the distribution of contaminant shifts is dominated by the brighter contaminating stars, with very faint contaminants unable to affect the flux-weighted position. Thus we choose $m = 5$, giving a flux ratio $F = 0.01$. For WISE in the Galactic plane, $l \simeq 120,\ b \simeq 0$, this gives a stellar density of $\simeq6\times10^4$ deg$^{-2}$ for $m_0 = 13$; a factor of 3 increase over Equation \ref{eq:expected3"contam}. The contamination levels themselves use for the area in question $R = 10"$, compared to the 0.5" used when calculating the false positive rate.

We find that inside one out of every four PSFs of stars of $W1 \simeq 13$ there will be a star of $13 \leq W1 \leq 15$. This increases to approximately one star of $15 \leq W1 \leq 17$ inside the PSF of every 13th magnitude WISE star. Naturally some of these objects will be deblended during the reduction process, meaning that these numbers are upper limits, but as Figure \ref{fig:crowding} demonstrates, not all of them are successfully recovered, meaning they must be buried within the brighter detections. 

\subsection{The Contamination Figure of Merit, Q}
\label{sec:figureofmerit}
The levels of contamination are dependent on the distribution of sources with magnitude and the size of the catalogue's PSF. To compare the contamination levels between catalogues requires a consistent metric.

Formally quantifying the stellar density requires fitting the number of stars per unit magnitude as a function of magnitude for the sky area in question. However, for a large fraction of the objects in the catalogue the contaminants that are perturbing their astrometry would be below the completeness limit of the catalogue, even outside of the bright star's PSF. This leads to the requirement that we extrapolate the number density of sources below the completeness limit. It is more straightforward to just consider the stellar density of the overall catalogue, and assume the extrapolation of the number density to faint magnitudes. 

We must decide on both a magnitude for which we will assess the contamination ($m_0$) and a maximum acceptable contamination level, before we can compare the contamination levels between catalogues. For $m_0$, we choose the median magnitude of the catalogue, which gives a lower bound to the contamination level of the fainter half of the catalogue. Additionally, we choose a contamination level of 33\%, the point at which a significant number of objects will be pertubed. These values then provide a baseline $Q$ value, which can then be compared to values calculated for specific catalogues.

The number of stars per unit area in the magnitude range from the median magnitude of the catalogue to five magnitudes fainter is approximately ten times that of the detected source density. We showed in Section \ref{sec:contamlevels} that contaminants more than five magnitudes fainter than the central object do not contribute to the overall perturbation, and we therefore limit ourselves to $m=5$. If we also wish to limit ourselves to 33\% of sources being contaminated, then

\begin{equation}
\int\limits_{m_0}^{m_0 + 5}\!N z^m\,\mathrm{d}m \times \pi R^2 = \frac{\mathrm{d}N}{\mathrm{d}A} \times \pi R^2 = 0.33.
\label{eq:contaminantfraction1}
\end{equation}
Substituting $R = 1.5\times$FWHM and $\frac{\mathrm{d}N}{\mathrm{d}A} = 10\times \frac{\mathrm{d}N}{\mathrm{d}A}_\mathrm{cat}$, where $\frac{\mathrm{d}N}{\mathrm{d}A}_\mathrm{cat}$ is the source catalogue density, we have

\begin{equation}
10 \times \frac{\mathrm{d}N}{\mathrm{d}A}_\mathrm{cat} \times \pi (1.5 \mathrm{FWHM})^2 = 0.33.
\label{eq:contaminantfraction2}
\end{equation}

This means that a 33\% contamination level of stars of the median magnitude is achieved when the contamination figure of merit

\begin{equation}
Q \equiv \frac{\mathrm{d}N}{\mathrm{d}A}_\mathrm{cat} \times \mathrm{FWHM}^2 = 0.005.
\label{eq:figureofmerit}
\end{equation}
It may be surprising that a catalogue where only a fraction of a percent of the sources might contain as contamination another source detected in the catalogue suffers from 33\% perturbation. Howevever, the 0.5\% result is simply the chance that a star above the completeness limit of the survey falls within a box with side length equal to the FWHM of the survey. The PSF length scale and, more importantly, the fact that stars are astrometrically perturbed by objects below the sensitivity of the survey both contribute to a much more significant contamination level. However, the $Q$ value is a useful tool for comparing surveys of different spatial resolutions and dynmical ranges.

Additionally, we can compare the number of objects affected both photometrically and astrometrically throughout the dynamical range of the catalogue. Towards the bright end of the catalogue, the number density of stars contaminating is relatively low. Here any stars affected will have accurate astrometric positions, and so the undetected contaminants will lead to large astrometric offsets compared to their uncertainties. However, the fraction of stars affected is sufficiently small that the contribution to the AUF from contaminated stars may be negligible. At the faint end of the catalogue the opposite is true, where the effective stellar density is very high and therefore the fraction of stars photometrically compromised is high. However, the SNR rapidly decreases towards the completeness limit of the survey and thus the influence of the contaminant stars is diminished, lost amidst the inherent uncertainty in measuring the position. Astrometrically the most affected part of the catalogue is between these two extremes, in the region where the stellar density is still high enough to have a large fraction of stars contaminated, but with accurate enough positions that the effects of contaminants are easily detectable.

\section{Surveys in Context and the Quoted-Core Distribution Uncertainty Relationship}
\label{sec:surveycontext}
While we have focused mostly on the WISE AUF, it is salient at this point to mention how this effect changes the distributions of other catalogues. Here we will briefly discuss three additional, complementary, large-scale surveys: two optical surveys, APASS and IPHAS, and the near-IR survey 2MASS. These catalogues are especially useful as they allow us to directly probe the effect of increasing stellar density and decreasing PSF scale length. We will also put WISE into a wider context.

We have shown evidence of a broadening of the AUFs relative to their assumed Gaussian positional uncertainties in Section \ref{sec:synthetictests}. As a consequence, we fit the AUFs for large sections of the Galaxy for each survey in one square degree divisions, giving relationships between the quoted and best-fit Rayleigh distribution uncertainties. The relationship between the quoted uncertainty and best fit Rayleigh distribution is
\begin{equation}
\sigma_\mathrm{core} = m\sigma_\mathrm{quoted} + c,
\label{eq:invsoutrelationship}
\end{equation}
with the core uncertainty such that the Rayleigh distribution best fits the smallest radial offsets of the given dataset, and the quoted uncertainty that as taken directly from their respective catalogues. We fit for some arbitrary offset $c$, but as expected the best fits have intercepts on the order $|c| \lesssim 0.05"$, resulting in effectively a scaling between the quoted and core uncertainties.

However, as detailed further in Section \ref{sec:summary}, while these broadened Gaussian uncertainties are useful, it must be cautioned that these empirical uncertainties do not necessarily allow for the selection of uncontaminated objects. Figure \ref{fig:fakeshifts} shows that there is significant overlap between the contaminated and uncontaminated distributions.

\subsection{APASS}
\label{sec:apasseffect}
As an all-sky survey bridging the gap between the Tycho2 and SDSS surveys \citep{Henden:2014aa}, APASS is a very important survey. However, it has a relatively large PSF, using a diameter of 15-20" for its aperture photometry, and large detector pixels ($\simeq3"$/pixel), leading to a significant fraction of contaminated stars and large wings in the APASS-\textit{Gaia} separation distribution. This is mitigated slightly by its reasonably bright completeness limit, effectively reducing the stellar density at its faint end, giving a contamination fraction on the order of tens of percent, or a $Q$ value of $3.4\times10^{-3}$.

APASS has very conservative astrometric uncertainties in DR9, requiring an empirical fit to any data being used in a probability-based matching process. In the Galactic plane (l $\simeq$ 120, b $\simeq$ 0) the core uncertainty is approximately 65\% of the quoted uncertainty, decreasingly dramatically towards the Galactic pole ($b\geq75$) where the core uncertainty is $\simeq$30\% of the quoted uncertainty.

\subsection{IPHAS}
\label{sec:iphaseffect}
IPHAS used the Isaac Newton Telescope on La Palma to conduct a relatively large scale, deep survey of a section of the Galactic plane. The median PSF FWHM of $\simeq1"$ combined with a 0.33" pixel scale \citep{Barentsen:2014tb} lead to a good ability to resolve sources even in crowded regions. In spite of this, IPHAS has a similar $Q$ value as APASS, at $4.4\times10^{-3}$, indicating a similar relative level of contamination at the two catalogues' respective median magnitudes. This results in a contamination fraction of 10-15\% at the faint end of the survey. Its much smaller PSF radius compared with APASS allows for a deeper survey at the same contamination level, or reduced contamination level at the same magnitude, as shown in Section \ref{sec:apassvsiphas}.

While the survey does not provide astrometric uncertainties for individual stellar sources, the high quality of the photometry means that there is good agreement between empirical distribution uncertainties and astrometric uncertainties calculated as the image FWHM divided by the photometric SNR, as per \citet{King:1983aa}.

\subsubsection{APASS vs IPHAS}
\label{sec:apassvsiphas}
As was seen in Sections \ref{sec:apasseffect} and \ref{sec:iphaseffect}, both optical catalogues have a similar $Q$ value - that is, the number of stars in an area the size of their PSF FWHM is similar. With the overlap in sky coverage and photometric bands, we can directly compare the separations of stars in common to both APASS and IPHAS with \textit{Gaia}. 

\begin{figure}
    \centering
    \includegraphics[width=\columnwidth]{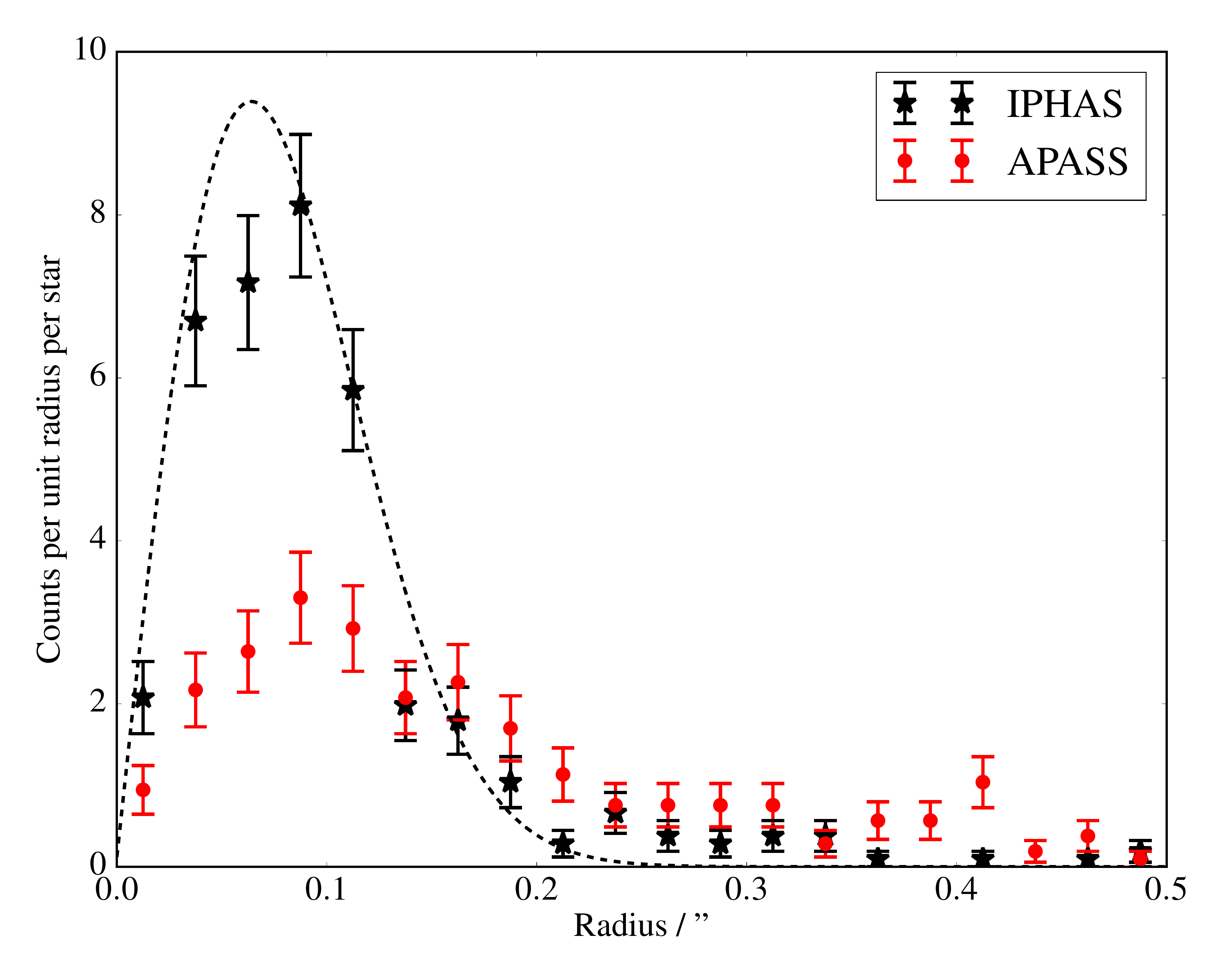}
    \caption{Figure comparing the effects of PSF resolution on the distribution of separations. Both IPHAS and APASS were matched to \textit{Gaia} and those in common where plotted for $\sigma_\mathrm{APASS} < 0.15"$, for IPHAS in black stars and APASS in red circles. The $\simeq1"$ FWHM of the IPHAS PSF gives contamination on the order of $\simeq5\%$ at an average of 15$^\mathrm{th}$ magnitude, whereas the $15-20"$ aperture used for APASS leads to much increased contamination, causing a much broadened distribution. Theoretical distribution of separations is shown as a dotted line for reference.}
    \label{fig:apassiphascomp}
\end{figure}

After matching both datasets to \textit{Gaia} for $120 \leq l \leq 125$, $0 \leq b \leq 5$, IPHAS and APASS stars which matched to the same \textit{Gaia} object were assumed to be themselves the same object. Stars were then selected with APASS astrometric uncertainties less than 0.15". Their separation distributions were then compared, as shown in Figure \ref{fig:apassiphascomp}.

The theoretical AUF matches the IPHAS distribution relatively well, with a small wing on the order of several percent, consistent with density contamination arguments. However, those same stars' positions are much more uncertain in APASS, caused in part by the differences in SNR, sky conditions etc., but additional broadening is caused by the vastly increased area subtended by stars on the sky in the APASS system.

As a consequence, we consider separately the magnitude at which a given catalogue will reach approximately 33\% contamination within its PSF. This value highlights the differences between APASS and IPHAS. The magnitude at which contamination of APASS sources up to five magnitudes fainter reaches 33\% is $B\simeq18.2$, which is approximately at the completeness limit of the survey in the uncrowded Galactic pole. However, the magnitude at which IPHAS suffers 33\% five magnitude fainter contamination is $r=23.4$, a few magnitudes fainter than its limiting magnitude of 20-21. This highlights the importance of spatial resolution on the contamination levels of photometric observations.

\subsection{2MASS}
\label{sec:2masseffect}
2MASS is frequently used to define the reference sky positions of those catalogues that came after it due to its all-sky completeness level (WISE and IPHAS both use it, for example), and therefore it is very important to understand the contamination levels that it suffers. However, it has a reasonably large PSF (FWHM $\simeq$ 2.5") and is a relatively faint (K$_s \simeq$ 16-17) survey. The contamination level rapidly increases with increasing magnitude and there are $\gtrsim0.8$ stars in every 2MASS PSF at its limiting magnitude in the Galactic plane. This results in $Q = 1.3\times10^{-2}$, or one in three contaminated stars with contaminants up to five magnitudes fainter at $J = 13.4$.

The quoted uncertainties match the core region of the distribution to within 10\%.

\subsection{WISE}
\label{sec:wiserelationship}
With its large, 10" PSF and high SNR leading to faint limiting magnitudes, WISE is especially susceptible to crowding, leading to, on average, one faint star inside every PSF of stars with $W1 \simeq 13$. WISE has an especially large $Q$ value, $\simeq6\times10^{-2}$. Its 33\% contamination level at the 1\% flux level is reached at a very bright magnitude as well, with one in three stars of $W1 = 9$ suffering from a star $9 \leq W1 \leq 14$ inside its PSF.

In the Galactic plane, we find that the core uncertainty is twice the quoted uncertainty, explained by the large fraction of contaminated stars. However, the Galactic poles suffer much less from contamination with its reduced stellar density. We therefore find that at $\sigma_\mathrm{quoted} \gtrsim 0.15"$ the quoted uncertainties fit the distributions with only minor broadening. Core uncertainties are only 10-15\% larger at these larger uncertainties, but below 0.15" the core uncertainty plateaus requiring a constant $\sigma$ to explain these brightest objects.

\subsection{\textit{Gaia}}
\label{sec:gaiacontext}
As a survey dedicated to astrometry, \textit{Gaia} has unparalleled precision in the positions of stars. Its PSF of 0.1" FWHM leads to a very small $Q$ value of $7.9\times10^{-5}$, 50 times better than any other catalogue used, or a limiting magnitude contamination of $\simeq0.1\%$. The magnitude at which contamination from stars five magnitudes fainter reaches 33\% is $G = 30.7$, far fainter than the completeness limit of the survey. From this we are confident in using \textit{Gaia} as the reference catalogue for quantifying the effects of contamination.

\section{How to Deal with Contaminated Astrometric Detections}
\label{sec:summary}
While the effect of unresolved objects inside stellar PSFs causing large wings to the probability distributions is explicit qualitatively, it is much more difficult to utilise it quantitatively. However, there are several ways to improve the matching process, depending on the specific requirements of the final catalogue of matches.

Two extremes of catalogue matching are the case where we must only return sources we can trust to not be contaminated or be false positives, and the case where we do not necessarily care whether our sources are contaminated, and are also willing to accept a large number of false positives. The decision may also be motivated by whether it is acceptable that matches have detections with fluxes that are compromised by a second star in their PSFs in one or both of the respective catalogues.

In either case, it should be noted that there will be some situations, such as with WISE-\textit{Gaia} matches, where one catalogue has a large PSF and the other has good spatial resolution, which will lead to a significant number of missed matches. These will be matches where one star contains within it as contamination a second object which is a separate entry in the opposite catalogue, which will lead to confusion in interpreting any results obtained. This will suggest that the faint \textit{Gaia} source has a corresponding WISE magnitude below WISE's completeness limit, which may not be the case in reality.

We also stress again that the contamination levels quoted here are upper limits, as active and passive deblending can help to resolve out overlapping objects, but note that this does not remove the effect entirely, as seen in the crowding out of stars (Figure \ref{fig:crowding}).

\subsection{Non-Contaminated Matches}
\label{sec:noncontammatch}
First, when the goal is to only match those stars which are definitely true matches, but now additionally are not significantly flux contaminated, it is advisable to cut proximity matches at a minimum of 3$\sigma_\mathrm{core}$. Equivalently, $\sigma_\mathrm{core}$ should be used as the uncertainty in the AUF when considering probability-based matches. 

We recommend examining sample distributions of proximity-matched separations. These should then be compared to their quoted uncertainty. If the quoted uncertainties are a good match to the empirical AUFs then use $\sigma_\mathrm{quoted}$, but otherwise make empirical corrections to fit the slightly broadened distributions to match as required. 

This will mostly capture the ``clean'' population, but will also increase the number of non-matches, as the AUF will not be sampling the extended tails of the contamination. This will potentially lead to the belief that the star was not detected in the opposing catalogue, with a cutoff radius that omits a large fraction of true matches. It will also still include some fraction of sources which are photometrically compromised, especially towards the fainter end of a given survey.

\subsection{Full Coverage Matches}
\label{sec:fullcovmatch}
The other extreme is the case where the goal is to achieve a large catalogue with as many matches as possible, in which the effect of false positives or contaminated fluxes is unimportant. 

In this case, the cutoff radius for a traditional nearest-neighbour match should be some multiple of the largest PSF FWHM between the two catalogues, typically 1.5-2 FWHMs. Alternatively, if a probability-based matching system is being used, then it is advisable to construct a set of empirical AUFs for each astrometric uncertainty slice in turn, which will include the wings of the distributions. 

These empirical functions are then used in place of $f$ as described by \citet{Sutherland:1992aa}, $g$ as per \citet{Naylor:2013aa}, $Q_{\chi^2}$ in \citet{Pineau:2017aa}, $LR_i$ of \citet{Rutledge:2000aa}, etc. These will increase the effective size of the area over which you can match between the catalogues, but will in turn increase the false match probability. Care should be taken when substituting any empirical functions into these probability-based matching methods, however, as any assumptions involving the use of Gaussian statistics (e.g., convolutions, mean positions, etc.) will no longer hold. 

We can take the WISE-\textit{Gaia} case to demonstrate the effects of an empirical AUF. To do so, we matched the two catalogues using a probability-based matching process (Wilson \& Naylor, in prep). The matching was done twice for two different astrometric PDFs. First, the AUFs used were purely Gaussian-based using $\sigma_\mathrm{quoted}$, and second, the WISE AUF was empirically constructed. When comparing the number of returned crossmatches, the Gaussian-based AUFs returned approximately half the pairs that the empirically constructed AUFs matched. Therefore, in crowded regions where the contamination of sources is high, probability-based matching using Gaussian statistics could result in as many as one in two true (albeit contaminated) counterparts being rejected as uncorrelated field objects.

\section{Conclusions}
\label{sec:conclusions}
We have presented an analysis of the distribution of WISE object positions with relation to \textit{Gaia} positions to determine their AUF, the probability density function of a catalogue's detected positions as a function of distance. We have found that the core of the distribution of separations can be fit with Gaussian statistics, although they require broadening, which we fit for empirically. However, there is an additional, significant, non-Gaussian tail to the distributions which is explained by flux contamination from fainter stars lying undetected within the PSF of the brighter star. In addition, we have discussed the contamination levels of APASS, IPHAS, and 2MASS.

We note that while we have focussed on the effects the contamination has on the measuring of individual positions, large tails are also seen in the distributions of proper motions (e.g., \citealp{Dong:2011aa}, \citealp{Feltzing:2002aa}, \citealp{Theissen:2016aa}). We suggest that the large tails seen in contaminated star positions could also propagate to explain the wings of these distributions, as proper motions are simply repeated astrometric measurements over a given time frame. 

We have focused on WISE in this work, as it is especially affected by this problem, because it reaches reasonably faint magnitudes in the infra-red and has a large PSF. However, it remains a problem for all catalogues, being an especially important consideration for the next generation of very deep ground-based surveys, such as LSST, with its predicted depth in the optical of $r\simeq$25 resulting in a theoretical $Q$ value of approximately $4\times10^{-2}$. This means that at fainter magnitudes most detected objects will be contaminated by one or more faint objects in their PSF. In comparion, \textit{Gaia} has a contamination level on the order of 0.1\%, due to its 0.1" FWHM PSF, meaning its positions should be robust against contamination.

\section*{Acknowledgements}
\label{sec:acknowledge}
TJW acknowledges support from an STFC Studentship. The authors wish to thank the reviewer for their general advice and specific help with clean \textit{Gaia} flags. This work has made use of the SciPy \citep{scipy}, NumPy \citep{numpy}, Matplotlib \citep{matplotlib}, and F2PY \citep{f2py} Python modules.

This research has made use of the APASS database, located at the AAVSO web site. Funding for APASS has been provided by the Robert Martin Ayers Sciences Fund. We would also like to thank the team personally for their support and feedback during the early stages of this work.

This paper makes use of data obtained as part of the INT Photometric H$\alpha$ Survey of the Northern Galactic Plane (IPHAS, www.iphas.org) carried out at the Isaac Newton Telescope (INT). The INT is operated on the island of La Palma by the Isaac Newton Group in the Spanish Observatorio del Roque de los Muchachos of the Instituto de Astrofisica de Canarias. All IPHAS data are processed by the Cambridge Astronomical Survey Unit, at the Institute of Astronomy in Cambridge. The bandmerged DR2 catalogue was assembled at the Centre for Astrophysics Research, University of Hertfordshire, supported by STFC grant ST/J001333/1. 

This publication makes use of data products from the Two Micron All Sky Survey, which is a joint project of the University of Massachusetts and the Infrared Processing and Analysis Center/California Institute of Technology, funded by the National Aeronautics and Space Administration and the National Science Foundation. 

This publication makes use of data products from the Wide-field Infrared Survey Explorer, which is a joint project of the University of California, Los Angeles, and the Jet Propulsion Laboratory/California Institute of Technology, funded by the National Aeronautics and Space Administration. 

This work has made use of data from the European Space Agency (ESA) mission \textit{Gaia} (\url{http://www.cosmos.esa.int/gaia}), processed by the \textit{Gaia} Data Processing and Analysis Consortium (DPAC, \url{http://www.cosmos.esa.int/web/gaia/dpac/consortium}). Funding for the DPAC has been provided by national institutions, in particular the institutions participating in the \textit{Gaia} Multilateral Agreement. 




\bibliographystyle{mnras}
\bibliography{mnrasastrometryproof.bib} 

\begin{thebibliography}{}
\makeatletter
\relax
\def\mn@urlcharsother{\let\do\@makeother \do\$\do\&\do\#\do\^\do\_\do\%\do\~}
\def\mn@doi{\begingroup\mn@urlcharsother \@ifnextchar [ {\mn@doi@}
  {\mn@doi@[]}}
\def\mn@doi@[#1]#2{\def\@tempa{#1}\ifx\@tempa\@empty \href
  {http://dx.doi.org/#2} {doi:#2}\else \href {http://dx.doi.org/#2} {#1}\fi
  \endgroup}
\def\mn@eprint#1#2{\mn@eprint@#1:#2::\@nil}
\def\mn@eprint@arXiv#1{\href {http://arxiv.org/abs/#1} {{\tt arXiv:#1}}}
\def\mn@eprint@dblp#1{\href {http://dblp.uni-trier.de/rec/bibtex/#1.xml}
  {dblp:#1}}
\def\mn@eprint@#1:#2:#3:#4\@nil{\def\@tempa {#1}\def\@tempb {#2}\def\@tempc
  {#3}\ifx \@tempc \@empty \let \@tempc \@tempb \let \@tempb \@tempa \fi \ifx
  \@tempb \@empty \def\@tempb {arXiv}\fi \@ifundefined
  {mn@eprint@\@tempb}{\@tempb:\@tempc}{\expandafter \expandafter \csname
  mn@eprint@\@tempb\endcsname \expandafter{\@tempc}}}

\bibitem[\protect\citeauthoryear{{Barentsen} et~al.,}{{Barentsen}
  et~al.}{2014}]{Barentsen:2014tb}
{Barentsen} G.,  et~al., 2014, \mn@doi [\mnras] {10.1093/mnras/stu1651}, 444,
  3230

\bibitem[\protect\citeauthoryear{{Budav{\'a}ri} \& {Szalay}}{{Budav{\'a}ri} \&
  {Szalay}}{2008}]{Budavari:2008aa}
{Budav{\'a}ri} T.,  {Szalay} A.~S.,  2008, \mn@doi [\apj] {10.1086/587156},
  679, 301

\bibitem[\protect\citeauthoryear{{Cutri} et~al.,}{{Cutri}
  et~al.}{2012}]{Cutri:2012aa}
{Cutri} R.~M.,  et~al., 2012, Technical report, {Explanatory Supplement to the
  WISE All-Sky Data Release Products}

\bibitem[\protect\citeauthoryear{{Dong}, {Gunn}, {Knapp}, {Rockosi}  \&
  {Blanton}}{{Dong} et~al.}{2011}]{Dong:2011aa}
{Dong} R.,  {Gunn} J.,  {Knapp} G.,  {Rockosi} C.,   {Blanton} M.,  2011,
  \mn@doi [\aj] {10.1088/0004-6256/142/4/116}, 142, 116

\bibitem[\protect\citeauthoryear{{Feltzing} \& {Johnson}}{{Feltzing} \&
  {Johnson}}{2002}]{Feltzing:2002aa}
{Feltzing} S.,  {Johnson} R.~A.,  2002, \mn@doi [\aap]
  {10.1051/0004-6361:20011771}, 385, 67

\bibitem[\protect\citeauthoryear{{Flesch} \& {Hardcastle}}{{Flesch} \&
  {Hardcastle}}{2004}]{Flesch:2004aa}
{Flesch} E.,  {Hardcastle} M.~J.,  2004, \mn@doi [\aap]
  {10.1051/0004-6361:20041076}, 427, 387

\bibitem[\protect\citeauthoryear{{Fleuren} et~al.,}{{Fleuren}
  et~al.}{2012}]{Fleuren:2012aa}
{Fleuren} S.,  et~al., 2012, \mn@doi [\mnras]
  {10.1111/j.1365-2966.2012.21048.x}, 423, 2407

\bibitem[\protect\citeauthoryear{{Gaia Collaboration}, {Brown}, {Vallenari},
  {Prusti}, {de Bruijne}, {Mignard}, {Drimmel}  \& {co-authors}}{{Gaia
  Collaboration} et~al.}{2016}]{Gaia-Collaboration:2016aa}
{Gaia Collaboration} {Brown} A.~G.~A.,  {Vallenari} A.,  {Prusti} T.,  {de
  Bruijne} J.,  {Mignard} F.,  {Drimmel} R.,   {co-authors} .,  2016, \aap,
  Special Gaia Volume

\bibitem[\protect\citeauthoryear{{Henden} \& {Munari}}{{Henden} \&
  {Munari}}{2014}]{Henden:2014aa}
{Henden} A.,  {Munari} U.,  2014, Contributions of the Astronomical Observatory
  Skalnate Pleso, 43, 518

\bibitem[\protect\citeauthoryear{{Herschel}}{{Herschel}}{1857}]{Herschel:1857aa}
{Herschel} J.~F.~W.,  1857, {Essays from the Edinburgh and Quarterly Reviews,
  with Addresses and Other Pieces}

\bibitem[\protect\citeauthoryear{{Hunter}}{{Hunter}}{2007}]{matplotlib}
{Hunter} J.~D.,  2007, Computing in Science \& Engineering, 9

\bibitem[\protect\citeauthoryear{{Jones}, {Oliphant}, {Peterson}
  et~al.}{{Jones} et~al.}{2001}]{scipy}
{Jones} E.,  {Oliphant} E.,  {Peterson} P.,   et~al., 2001, {SciPy}: Open
  Source Scientific Tools for {Python}

\bibitem[\protect\citeauthoryear{{King}}{{King}}{1983}]{King:1983aa}
{King} I.~R.,  1983, \mn@doi [\pasp] {10.1086/131139}, 95, 163

\bibitem[\protect\citeauthoryear{{Krawczyk}, {Richards}, {Mehta}, {Vogeley},
  {Gallagher}, {Leighly}, {Ross}  \& {Schneider}}{{Krawczyk}
  et~al.}{2013}]{Krawczyk:2013aa}
{Krawczyk} C.~M.,  {Richards} G.~T.,  {Mehta} S.~S.,  {Vogeley} M.~S.,
  {Gallagher} S.~C.,  {Leighly} K.~M.,  {Ross} N.~P.,   {Schneider} D.~P.,
  2013, \mn@doi [\apjs] {10.1088/0067-0049/206/1/4}, 206, 4

\bibitem[\protect\citeauthoryear{Line, Webster, Pindor, Mitchell  \&
  Trott}{Line et~al.}{2017}]{2017PASA...34....3L}
Line J. L.~B.,  Webster R.~L.,  Pindor B.,  Mitchell D.~A.,   Trott C.~M.,
  2017, \mn@doi [\pasa] {10.1017/pasa.2016.58}, 34, e003

\bibitem[\protect\citeauthoryear{Michalik, Lindegren  \& Hobbs}{Michalik
  et~al.}{2015}]{2015A&A...574A.115M}
Michalik D.,  Lindegren L.,   Hobbs D.,  2015, \mn@doi [\aap]
  {10.1051/0004-6361/201425310}, 574, A115

\bibitem[\protect\citeauthoryear{{Munari} et~al.,}{{Munari}
  et~al.}{2014}]{Munari:2014aa}
{Munari} U.,  et~al., 2014, \mn@doi [\aj] {10.1088/0004-6256/148/5/81}, 148, 81

\bibitem[\protect\citeauthoryear{{Naylor}, {Broos}  \& {Feigelson}}{{Naylor}
  et~al.}{2013}]{Naylor:2013aa}
{Naylor} T.,  {Broos} P.~S.,   {Feigelson} E.~D.,  2013, \mn@doi [\apjs]
  {10.1088/0067-0049/209/2/30}, 209, 30

\bibitem[\protect\citeauthoryear{{Peterson}}{{Peterson}}{2009}]{f2py}
{Peterson} P.,  2009, International Journal of Computational Science and
  Engineering, 4

\bibitem[\protect\citeauthoryear{{Pineau} et~al.,}{{Pineau}
  et~al.}{2017}]{Pineau:2017aa}
{Pineau} F.-X.,  et~al., 2017, \mn@doi [\aap] {10.1051/0004-6361/201629219},
  597, A89

\bibitem[\protect\citeauthoryear{{Rutledge}, {Brunner}, {Prince}  \&
  {Lonsdale}}{{Rutledge} et~al.}{2000}]{Rutledge:2000aa}
{Rutledge} R.~E.,  {Brunner} R.~J.,  {Prince} T.~A.,   {Lonsdale} C.,  2000,
  \mn@doi [\apjs] {10.1086/317363}, 131, 335

\bibitem[\protect\citeauthoryear{{Skrutskie} et~al.,}{{Skrutskie}
  et~al.}{2006}]{Skrutskie:2006um}
{Skrutskie} M.~F.,  et~al., 2006, \mn@doi [\aj] {10.1086/498708}, 131, 1163

\bibitem[\protect\citeauthoryear{{Sutherland} \& {Saunders}}{{Sutherland} \&
  {Saunders}}{1992}]{Sutherland:1992aa}
{Sutherland} W.,  {Saunders} W.,  1992, \mn@doi [\mnras]
  {10.1093/mnras/259.3.413}, 259, 413

\bibitem[\protect\citeauthoryear{{Theissen}, {West}  \& {Dhital}}{{Theissen}
  et~al.}{2016}]{Theissen:2016aa}
{Theissen} C.~A.,  {West} A.~A.,   {Dhital} S.,  2016, \mn@doi [\aj]
  {10.3847/0004-6256/151/2/41}, 151, 41

\bibitem[\protect\citeauthoryear{{Wright} et~al.,}{{Wright}
  et~al.}{2010}]{Wright:2010aa}
{Wright} E.~L.,  et~al., 2010, \mn@doi [\aj] {10.1088/0004-6256/140/6/1868},
  140, 1868

\bibitem[\protect\citeauthoryear{{van der Walt}, {Colbert}  \&
  {Varoquaux}}{{van der Walt} et~al.}{2011}]{numpy}
{van der Walt} S.,  {Colbert} S.~C.,   {Varoquaux} G.,  2011, Computing in
  Science \& Engineering, 13

\makeatother
\end{thebibliography}

\bsp	
\label{lastpage}
\end{document}